\documentclass[11pt]{article}
\usepackage{epsf} 
\usepackage{amsmath}
\usepackage{amssymb}
\usepackage{epsfig}
\usepackage{latexsym}
\usepackage{amsfonts}
\usepackage{graphicx}%
\usepackage{varioref}
\usepackage{ifthen}
\begin{document}
\parindent 0mm 
\setlength{\parskip}{\baselineskip} 
\thispagestyle{empty}
\pagenumbering{arabic} 
\setcounter{page}{0}
\mbox{ }
\rightline{UCT-TP-278/2010}
\newline
\newline
\rightline{February  2010}
\newline

\noindent
\begin{center}
\Large \textbf{Electromagnetic Form Factors of Hadrons in Dual-Large $N_c$ QCD $^1$}
\end{center}
{\LARGE \footnote{{\LARGE {\footnotesize Invited talk at the XII Mexican Workshop on Particles \& Fields, Mazatlan, November 2009. To be published in American Institute of Physics Conference Proceedings Series.
Work supported in part by NRF (South Africa). }}}}

\begin{center}
C. A. Dominguez
\end{center}

\begin{center}
Centre for Theoretical Physics and Astrophysics\\
University of Cape Town, Rondebosch 7700, South Africa, and Department of Physics, Stellenbosch University, Stellenbosch 7600, South Africa
\end{center}

\begin{center}
\textbf{Abstract}
\end{center}
\noindent
In this talk,  results are presented of determinations of electromagnetic form factors of hadrons (pion, proton, and $\Delta(1236)$)  in the framework of  Dual-Large $N_c$ QCD (Dual-$QCD_\infty$). This framework improves considerably tree-level VMD results by incorporating an infinite number of zero-width resonances, with masses and couplings fixed by the dual-resonance (Veneziano-type) model.
\newpage
\section{Introduction}
In its original formulation \cite{SAKU}, Vector Meson Dominance (VMD) is an effective tree-level model based on  $\gamma-\rho^0$ conversion. If used to compute electromagnetic form factors, it can very roughly account for the pion form factor data in the space-like region, and with some modifications (unitarization),  also in the time-like region around the rho-meson peak. However, for non-zero spin hadrons such as nucleons and $\Delta(1236)$, VMD is in serious disagreement with the observed $q^2$ fall-off of these form factors. It has been generally believed that the reason for this discrepancy is that the radial excitations of the $\rho$-meson are not taken into account in naive VMD. In fact, an infinite sum of monopole terms with suitable coefficients can lead to a form factor with an asymptotic behavior different from a monopole.
An attempt to remedy this situation was made long ago by incorporating radial excitations of the rho-meson into VMD, i.e. Extended VMD \cite{EVMD}. At the time, though, there was no known renormalizable QFT to support this approach. Today, we know that in the limit of an infinite number of colors, QCD is solvable  and leads to a hadronic spectrum consisting of an infinite number of zero-width states \cite{QCDINF}. However, the masses and couplings of these states remain unspecified, so that models are needed to fix these parameters. An attractive and highly economical candidate  is Dual-$QCD_\infty$ \cite{CAD1}-\cite{CAD3}, {\bf inspired} in the Dual Resonance Model for scattering amplitudes of Veneziano \cite{VEN}, the precursor of string theory. It is very important to stress the word {\bf inspired}, as Dual-$QCD_\infty$ does not share any of the unwanted features of the original Veneziano model for n-point functions ($n\geq 4$), such as lack of unitarity, unphysical particles (tachyons) in the spectrum, etc. These features simply do not emerge for three-point functions.  Another aspect of Dual-$QCD_\infty$ which needs to be stressed to avoid misunderstandings is that it is not intended to be an expansion in powers of $1/N_c$. In fact, $N_c$ is taken to be infinite from the start, as this is the limit in which QCD is solvable and leads to the hadronic spectrum mentioned above. Unitarization can subsequently be performed by shifting the poles from the real axis into the second Riemann sheet in the complex energy (squared) plane. This induces corrections to form factors of order $\cal{O}$$(\Gamma/M \simeq 10\%)$.\\
\section{Dual-$QCD_\infty$}
\begin{figure}
\begin{center}
  \includegraphics[height=.3\textheight]{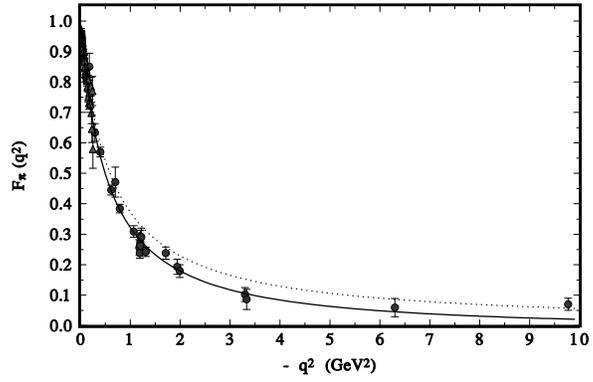}
  \caption{Pion form factor in Dual-$QCD_\infty$ (solid curve), and VMD (broken curve).}
  \label{fig:figure1}
  \end{center}
\end{figure}
\begin{figure}
\begin{center}
  \includegraphics[height=.25\textheight]{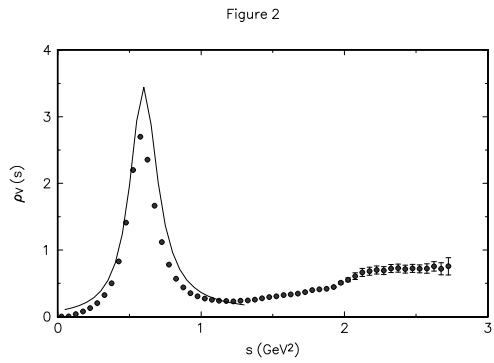}
  \caption{Pion form factor in Dual-$QCD_\infty$ in the time-like region.}
\label{fig:figure2}
\end{center}
\end{figure}
\begin{figure}
\begin{center}
  \includegraphics[height=.3\textheight]{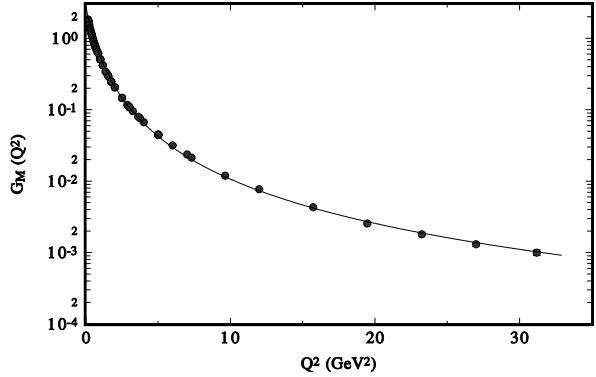}
  \caption{Sachs magnetic form factor of the proton, $G_M(Q^2)$. Data is from the reanalysis of \cite{NDATA}, and $Q^2\equiv - q^2$.}
\label{fig:figure3}
\end{center}
\end{figure}
In $QCD_\infty$, a typical form factor has the generic form
\begin{equation}
F(s) = \sum_{n=0}^{\infty}
\frac{C_{n}}{(M_{n}^{2} -s)} \;,
\end{equation}
where $s \equiv q^2$ is the momentum transfer squared, and  the masses $M_n$, and the couplings $C_n$ remain unspecified. In Dual-$QCD_{\infty}$ they are given by \cite{CAD1}
\begin{equation}
C_{n} = \frac{\Gamma(\beta-1/2)}{\alpha' \sqrt{\pi}} \; \frac{(-1)^n}
{\Gamma(n+1)} \;
\frac{1}{\Gamma(\beta-1-n)} \;, 
\end{equation}
where $\beta$ is a free parameter, and the string tension $\alpha'$ is $\alpha' = 1/2 M_{\rho}{^2}$,  as it enters the rho-meson Regge trajectory
$\alpha_{\rho}(s) = 1 + \alpha ' (s-M_{\rho}^{2})$. The mass spectrum is chosen as
$M_{n}^{2} = M_{\rho}^{2} (1 + 2 n)$. This simple formula correctly predicts the first few radial excitations. Other, e.g. non-linear mass formulas could be used, but this hardly changes the results in the space-like region, and only affects the time-like region behavior for very large $q^2$. With these choices the form factor becomes an Euler Beta-function, i.e.
\begin{eqnarray}
F(s) &=& \frac{\Gamma(\beta-1/2)}{\sqrt{\pi}} \; \sum_{n=0}^{\infty}\;
\frac{(-1)^{n}}{\Gamma(n+1)} \; \frac{1}{\Gamma(\beta-1-n)}\frac{1}
{[n+1-\alpha_\rho(s)]} \nonumber\\ [.5cm]
& = &
\frac{1}{\sqrt{\pi}} \; \frac{\Gamma (\beta-1/2)}{\Gamma(\beta-1)}   \;\;
B(\beta - 1,\; 1/2 - \alpha' s)\;,
\end{eqnarray}
where $B(x,y) = \Gamma(x) \Gamma(y)/\Gamma(x+y)$. The form factor exhibits asymptotic power behavior in the space-like region, i.e.
\begin{equation}
\lim_{s \rightarrow - \infty} F(s) = (- \alpha' \;s)^{(1-\beta)}\;,
\end{equation}
from which one identifies the free parameter $\beta$ as controlling this asymptotic behavior. Notice that while each term in Eq.(3) is of the monopole form, the result is not necessarily of this form because it involves a sum over an infinite number of states. The exception occurs for integer values of $\beta$, which leads to a finite sum.
The imaginary part of the form factor Eq.(3) is
\begin{equation}
Im \; F(s) = \frac{\Gamma(\beta-1/2)}{\alpha' \sqrt{\pi}} \;
\sum_{n=0}^{\infty} \; \frac{(-1)^{n}}{\Gamma(n+1)} 
 \frac{1}{\Gamma(\beta-1-n)} \; \pi \; \delta(M_n^2-s) \;.
\end{equation}
Unitarization can be performed by shifting the poles from the real axis in the complex s-plane. The simplest model is the Breit-Wigner form
\begin{equation}
\pi \delta(M_{n}^{2} -s) \rightarrow \frac{\Gamma_{n} M_{n}}
{[(M_{n}^{2} - s)^{2} + \Gamma_{n}^{2} M_{n}^{2}]} \;,
\end{equation}
where one expects $\Gamma_n$ to grow with $M_n$. Other, more refined choices, are certainly possible, e.g. the Gounaris-Sakurai  form in which the width is momentum transfer dependent.\\ 
\begin{figure}
\begin{center}
  \includegraphics[height=.25\textheight]{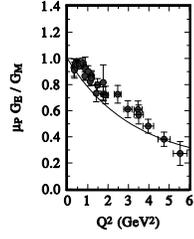}
  \caption{Ratio of electric to magnetic Sachs form factors of the proton. Data is from the reanalysis of \cite{NDATA}, and $Q^2\equiv - q^2$.}
\label{fig:figure4}
\end{center}
\end{figure}
\begin{figure}
\begin{center}
  \includegraphics[height=.3\textheight]{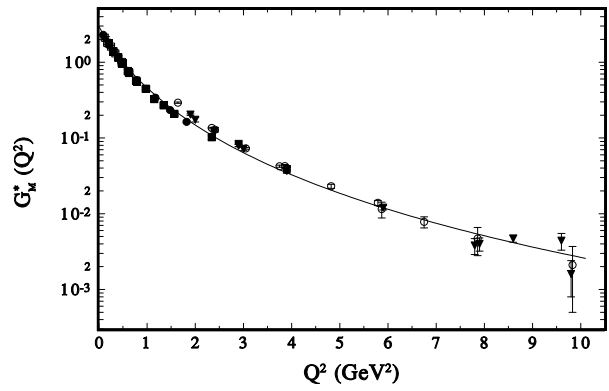}
  \caption{$\Delta(1236)$ magnetic form factor in Dual-$QCD_\infty$ (solid line).}
\label{fig:figure5}
\end{center}
\end{figure}
Results in this framework are shown, together with the data, in Fig. 1 and Fig.2 for the pion (data from \cite{DATAPI}), Figs. 3 and 4 for the proton, and Fig. 5  for the $\Delta(1236)$. The time-like pion form factor was obtained using Eq.(6); in spite of the simplicity of the model the agreement is reasonable at and around the $\rho$-peak. In the case of the proton, Eq.(3) is used for the Dirac and Pauli form factors, $F_1(q^2)$ and $F_2(q^2)$, as these have the correct analyticity properties. The fits have not been made  to the raw data, but rather to the  data base as corrected in \cite{NDATA}. These corrections take into account the discrepancies between unpolarized (SLAC) and polarized (JLAB) experiments. For the $\Delta(1236)$, the three so called Scadron form factors $G^*_{M,E,C}(q^2)$ were fitted using Eq.(3), and data on $G^*_M(q^2)$, and the two ratios between $G^*_{E,C}(q^2)$ and $G^*_M(q^2)$ \cite{DATA1}-\cite{DATA2}. The value of the free parameter $\beta$ in the form factor, Eq. (3), which determines its asymptotic behavior, is as follows: for the pion, $\beta_\pi = 2.3$, for $F_1$ and $F_2$ of the proton, $\beta_1 = 2.95 - 3.03$, and $\beta_2=4.13 - 4.20$, respectively, and for $G^*_M$, $G^*_E$, $G^*_C$ of the $\Delta(1236)$, $\beta^*_M = 4.6 - 4.8$, $\beta^*_E \simeq \beta^*_M$, and $\beta^*_C = 6.0 - 6.2$. Taking the middle values of these numbers, the asymptotic behavior in the space-like region of these form factors is approximately as follows: $F_\pi \sim (- q^2)^{- 1.3}$,
$F_1 \sim (- q^2)^{- 2.0}$, $F_2 \sim (- q^2)^{- 3.2}$, $G^*_M \sim G^*_E \sim (- q^2)^{- 3.7}$, and $G^*_C \sim (- q^2)^{- 5.1}$. 
\section{Conclusions}
Dual-$QCD_\infty$ accounts quite successfully for the space-like behavior of the pion, the proton, and the $\Delta(1236)$ electromagnetic form factors. A very simple unitarization procedure leads to a pion form factor in the time-like region in reasonable agreement with data (more sophisticated procedures improve considerably this agreement \cite{BRUCH}). Nucleon form factors in this framework and  in the time-like region are currently under investigation \cite{CAD4}
\section{Acknowledgments}
This talk is based on work done by the author, and in collaboration with  R. R\"{o}ntsch,  and T. Thapedi . Work supported in part by the NRF (South Africa). The author thanks the organizers for  an interesting and fruitful workshop.

\end{document}